\documentclass{article}

\usepackage[english]{babel}
\usepackage[backend=biber, style = nature, citestyle = nature, sorting=none]{biblatex}
\usepackage{amssymb}
\usepackage{bbding}

\usepackage[letterpaper,top=2cm,bottom=2cm,left=3cm,right=3cm,marginparwidth=1.75cm]{geometry}

\usepackage{amsmath}
\usepackage{graphicx}

\usepackage[colorlinks=true, allcolors=blue]{hyperref}
\usepackage{subcaption,booktabs}
\usepackage{xcolor,colortbl}
\definecolor{lavender}{rgb}{0.9, 0.9, 0.98}

\usepackage{siunitx}
\bibliography{main.bib}

\title{Inverse Designed WS$_2$ Planar Chiral Metasurface with Geometric Phase}

\usepackage{authblk}

\author[1]{Jaegang Jo}
\author[2]{Sangbin Lee}
\author[1]{Munseong Bae}
\author[3]{Damian Nelson}
\author[3,4]{Kenneth B. Crozier}
\author[5]{Nanfang Yu}
\author[1,2,a,*,\Envelope$^1$]{Haejun Chung}
\author[4,$*$,\Envelope$^2$]{Sejeong Kim}

\affil[1]{Department of Electronic Engineering, Hanyang University, Seoul, 04763, Republic of Korea}
\affil[2]{Graduate School of Artificial Intelligence Semiconductor, Hanyang University, Seoul, 04763, Republic of Korea}
\affil[3]{School of Physics, University of Melbourne, Victoria, 3010, Australia}
\affil[4]{Department of Electrical Engineering, Faculty of Engineering and Information Technology, University of Melbourne, Melbourne 3000, Australia}
\affil[5]{Department of Applied Physics and Applied Mathematics, Columbia University, New York, NY 10031, USA}
\affil[$*$]{These authors are corresponding authors.}

\begin{document}

\maketitle

\begin{center}
    \Envelope$^1$ haejun@hanyang.ac.kr
    \Envelope$^2$ sejeong.kim@unimelb.edu.au
\end{center}

\vspace{1pc}

\begin{abstract}
Increasing attention is being paid to chiral metasurfaces  due to their ability to selectively manipulate right-hand circularly polarized light or left-hand circularly polarized light. The thin nature of metasurfaces, however, poses a challenge in creating a device with effective phase modulation. Plasmonic chiral metasurfaces have attempted to address this issue by increasing light-matter interaction, but they suffer from metallic loss. Dielectric metasurfaces made from high index materials enable phase modulation while being thin. Very few materials, however, have high refractive index and low loss at visible wavelengths. Recently, some 2D materials have been shown to exhibit high refractive index and low loss in the visible wavelengths, positioning them as promising platform for meta-optics. This study introduces and details a planar chiral metasurface with geometric phase composed of WS$_2$ meta-units. By employing adjoint optimization techniques, we achieved broadband circular dichroism.
\end{abstract}
\vspace{2pc}
\noindent{\it Keywords}: Chiral metasurface, transition metal dichalcogenide (TMD), WS$_2$, dielectric metasurface, geometric phase, inverse design, adjoint optimization

%
%
%

\section{Introduction}
Metasurfaces, which consist of ultrathin metallic or dielectric nanostructures with dimensions smaller than the wavelength of light, have gained attention for their ability to manipulate optical phase, amplitude and polarization over the wavefront with subwavelength resolution.~\cite{malek2022multifunctional, overvig2019dielectric, yu2011light, jung2021metasurface} They have found numerous applications, including ultra-thin meta-lenses~\cite{kim2023scalable,chung2020high, khorasaninejad2016metalenses}, holograms~\cite{malek2022multifunctional, kim2021geometric, yoon2017pragmatic, so2023multicolor,lee2018metasurface, wen2021light}, and optical vortex beam generators~\cite{ren2020complex,yu2011light,bae2023inverse, chen2017digitalized, white2022inverse}. Geometric metasurfaces utilizing Pancharatnam–Berry phase have attracted considerable attention due to their broadband and efficient manipulation of optical phase.~\cite{jisha2021geometric, overvig2019dielectric, kim2023scalable, jung2021metasurface, xie2021generalized} The geometric phase is achieved by rotating a unit cell or meta-unit of metasurfaces, which converts circularly polarized incident light into transmitted light wave with the opposite handedness. This spin-switching capability enables a range of applications, including chiral-sensitive imaging~\cite{basiri2019nature, khorasaninejad2016multispectral} and display~\cite{kim2021geometric, chen20193d, lee2018metasurface, li2016multicolor}, and optical spin-orbit coupling~\cite{bliokh2015spin, chen2017digitalized}. However, conventional geometric metasurfaces, featuring in-plane mirror symmetry in their meta-units, are limited in their capability to independently modulate the two circularly polarized states. To address this issue, chiral metasurfaces have been proposed, with their broken in-plane mirror symmetry allowing independent control of the two circular polarization states of the transmitted light.~\cite{ma2018all, semnani2020spin, wang2022dynamic, naeem2023dynamic, chen2018spin} Such metasurfaces can be designed, for example, to maximize the transmission of circularly polarized light with a particular handedness while simultaneously blocking that with the opposite handedness.

Chiral metasurfaces can be designed with two types of materials: metals and dielectrics. While plasmonic chiral metasurfaces have demonstrated significant circular dichroism (CD) in the visible range, they suffer from significant ohmic losses.~\cite{chen2018spin, ji2021chirality, semnani2020spin, tang2017chiral}
Chiral metasurfaces made of dielectric material display negligible absorption; however, they often exhibit low CD or require considerable thickness to attain sufficient light-matter interactions for effective phase manipulation.~\cite{ma2018all, naeem2023dynamic, wang2022dynamic, shi2022planar} Therefore, it is challenging to create a thin, planar device that shows both high CD and low absorption in the visible.

In order to achieve a high level of CD in the visible range, we propose a planar chiral metasurface based on WS$_2$. WS$_2$ belongs to the family of Transition Metal Dichalcogenides (TMDs) and has attracted considerable attention in the nanophotonics and metasurfaces community due to its high refractive index and strong optical anisotropy.~\cite{kim2023all, zhang2020ultrathin, vyshnevyy2023van, choudhury2018material, verre2019transition} The high refractive index of WS$_2$, larger than 4 in the visible range, allows for a strong interaction between light and the material~\cite{shim2021fundamental}, which increases the polarization conversion efficiency and therefore decreases the required material thickness. 

In this work, we use an adjoint optimization method~\cite{lin2019topology, nelson2023inverse, chung2020high, bae2023inverse} to create planar chiral metasurfaces made of WS$_2$. This design technique optimizes the structure by using the gradient of the figure of merit (FOM) with respect to the design parameters through direct and adjoint full-wave simulations. The adjoint optimization enables the realization of a freeform structure that maximizes the FOM while satisfying the design constraints, resulting in a significant performance improvement over conventional heuristic design approaches. The WS$_2$ chiral metasurface in this work exhibits a high extinction ratio (peak value up to 19.6dB) in the visible range. Additionally, to verify the performance of the geometric phase manipulation of our designed chiral resonator, we analyze the characteristics of metasurfaces composed of chiral resonators with various rotation angles. These metasurfaces maintain consistently high CDs ($>$0.5) and transmittance ($>$0.57) at all rotation angles. 

\section{Results and Discussion}

\begin{figure}
    \centering
    \includegraphics[width=1.0\linewidth]{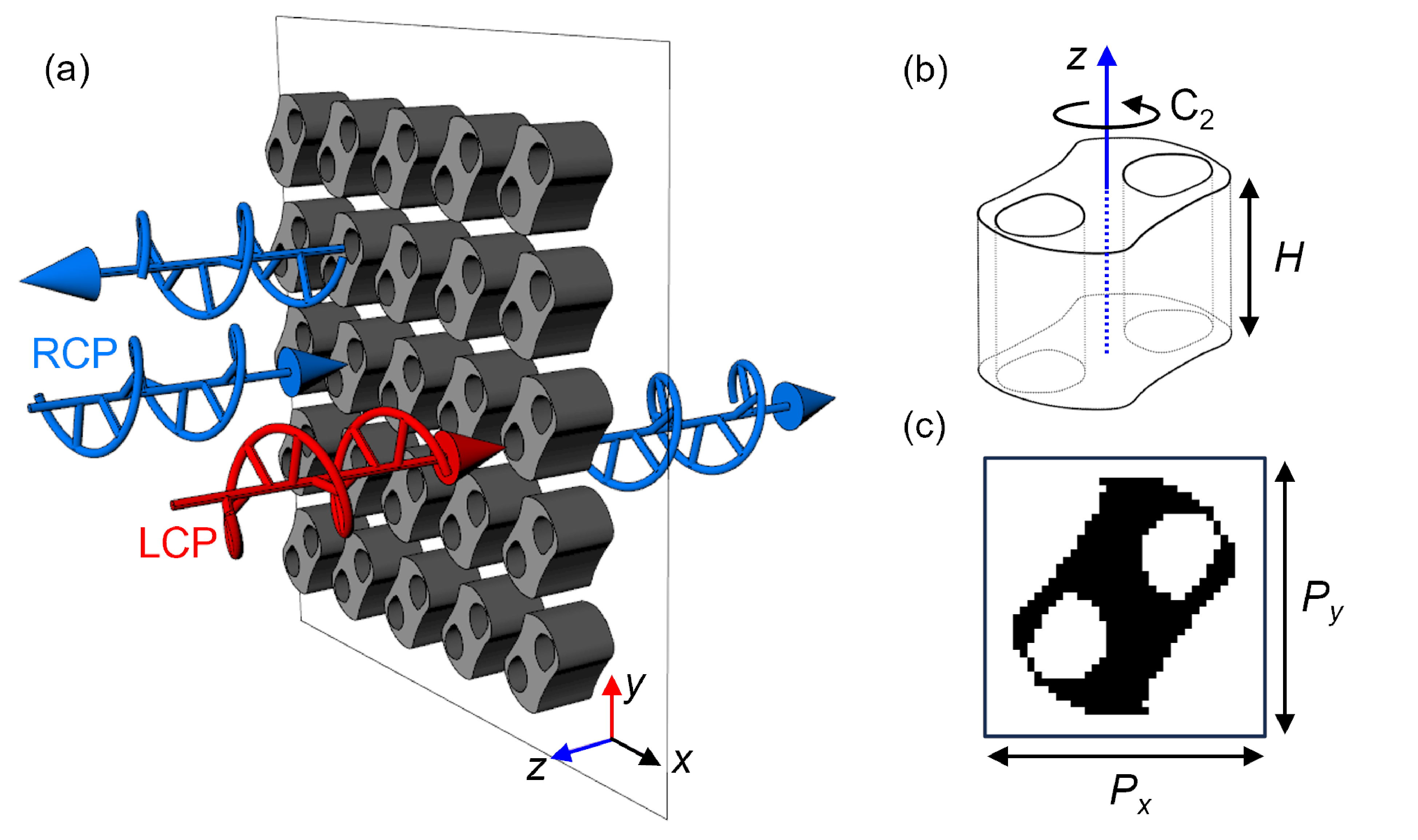}
    \caption{(a) Schematic illustration of the inversely-designed chiral metasurface optimized to reflect right-hand circularly polarized light (RCP, represented with blue colors) while maintaining the polarization and convert left-hand circularly polarized light(LCP, represented with red) to RCP. The metasurface consists of a periodic array of WS$_2$ chiral resonators on a glass substrate. (b) Illustration of a single WS$_2$ chiral resonator. The resonator has a two-fold rotational symmetry (C$_2$) along the $z$-axis, and the thickness ($H$) is 250 nm. (c)  Top-view of the pixelated chiral resonator in simulations. The black and white pixels represent WS$_2$ and air, respectively, and the single pixel dimension is $10\times10\times10$  nm$^3$. The periodicities of an array of the chiral resonators, $P_x$ and $P_y$, are 380 nm.}
    \label{fig:fig1}
\end{figure}

Figure~\ref{fig:fig1}(a) illustrates the chiral metasurface and its function, i.e., reflecting a right-hand circularly polarized light (RCP) wave at normal incidence while converting left-hand circularly polarized light (LCP) into an RCP wave in transmission. In this work, WS$_2$ is chosen for its high refractive index and relatively low absorption in the visible. The WS$_2$ metasurface is located on a glass substrate with a refractive index of 1.5. Simulations in this work used wavelength-dependent in-plane and out-of-plane refractive indices of WS$_2$, which are imported from the literature~\cite{vyshnevyy2023van}. 
The relationship between the incident and transmitted circularly polarized light~\ref{eqn:eq1} can be written using the Jones matrix calculus in the following form:
\begin{equation}
\begin{pmatrix}E_{tl}\\E_{tr}\end{pmatrix}
    =\begin{pmatrix}t_{ll} & e^{-i2\theta} t_{lr} \\
        e^{i2\theta}t_{rl} & t_{rr}
    \end{pmatrix}
    \begin{pmatrix}E_{il}\\E_{ir}\end{pmatrix} 
    \label{eqn:eq1}
\end{equation} 
Here, the subscripts $l$ and $r$ represent LCP and RCP, respectively while $i$ and $t$ denote incident and transmitted light, respectively. To obtain perfect conversion between RCP and LCP light for realizing metasurface based on the Pancharatnam-Berry phase, the diagonal elements of the matrix, $t_{ll}$ and $t_{rr}$, should be zero. Assuming the metasurface can minimize these terms, subsequently, the matrix is simplified, yielding two equations: $E_{tl} = e^{-i2\theta}t_{lr}E_{ir}$ and $E_{tr} = e^{i2\theta}t_{rl}E_{il}$, demonstrating the conversion of handedness. To create a metasurface that produces conversion only for LCP, we introduce a design goal of minimizing the conversion of RCP to LCP, i.e., $t_{lr} = 0$. 

Each chiral structure exhibits C$_2$ symmetry as shown in Fig.~\ref{fig:fig1}(b), a constraint we impose for structure optimization. This is because the generated phase modulation covers the full 2$\pi$ range when the change in geometrical rotation, $\theta$, corresponds to $\pi$. A range of different heights of the resonator is tested, and 250 nm demonstrates the highest figure of merit (FoM). In the 3D simulation domain, the structure is pixelated as shown in Fig. 1(c), where each pixel has a volume of $10\times10\times10$ nm$^3$. A square lattice array is used with an array periodicity of 380 nm in both the $x$ and $y$ directions. 

\begin{figure}
    \centering
    \includegraphics[width=1.0\linewidth]{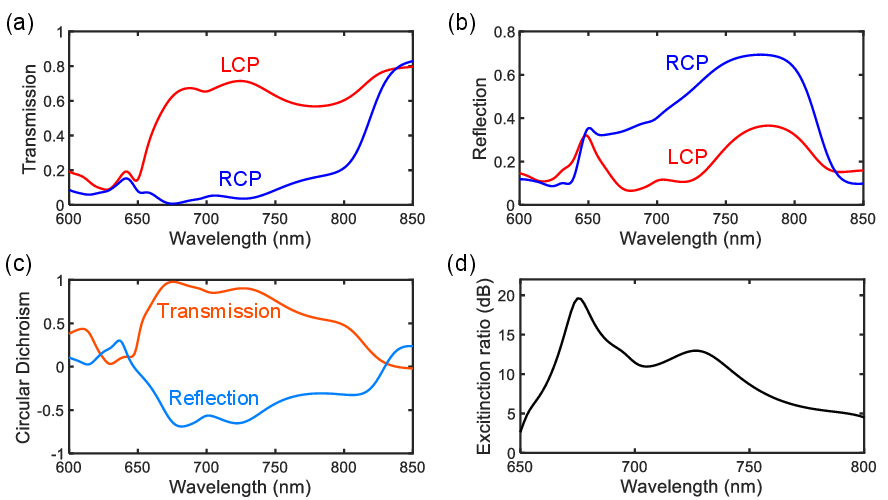}
    \caption{(a) Transmission spectra when LCP (red) and RCP (blue) are incident onto the inversely-designed chiral resonator array. (b) Reflection spectra of LCP and RCP incidences. (c) Circular dichroism, the difference in light intensity between LCP and RCP, is obtained from the transmission and reflection data in (a) and (b). (d) Calculated extinction ratio, i.e., ration of transmittance under RCP and LCP incidence, of the metasurface, with a peak reaching 19.6dB at $\lambda$=675 nm.}
    \label{fig:fig2}
\end{figure}

Figure~\ref{fig:fig2}(a) shows the transmission and reflection spectra occurring with LCP and RCP illumination of the inverse-designed chiral metasurface. The transmission monitor in this simulation measures the power of the transmitted light and therefore does not distinguish the state of polarization. It is noteworthy to mention that the dichroism is significant across a wide range of the wavelengths, unlike most previous works where high dichroism is usually limited to a narrow bandwidth.~\cite{ma2018all, shi2022planar, semnani2020spin, tang2010optical} Substantial light intensity contrast is also observed in reflection spectra as shown in Fig.~\ref{fig:fig2}(b) with RCP being reflected more than the LCP. Figure~\ref{fig:fig2}(c) shows the circular dichroism (CD) of the metasurface in transmission (red line) and reflection (blue line). The transmission CD is defined as ($T_L-T_R$)/($T_L+T_R$), where $T_L$ and $T_R$ are the transmissions under the LCP and RCP incidences, respectively. Likewise, the reflection CD is defined as ($R_L-R_R$)/($R_L+R_R$). The CD reaches a maximum value of close to unity at $\lambda=$675 nm, and exceed 0.5 across spectral window that is $\sim$150 nm wide. Figure 2 (d) shows the extinction ratio, defined as $T_L$/$T_R$, with the peak value reaching 19.6dB at 675 nm. This value is significantly high compared to those of reported chiral metasurfaces.~\cite{ma2018all, shi2022planar, tang2017chiral}.

The very high extinction ratio of the metasurface is enabled by adjoint-optimization-based inverse design with a customized figure of merit (FOM). The FOM ($\mathcal{F}$) of adjoint optimization can be expressed as an inner product between a forward field $\textbf{E}$ and a target field $\textbf{E}_d$.
\begin{equation}
\mathcal{F} = \int_\mathcal{M} |\textbf{E}(\textbf{x}) \cdot \textbf{E}_d(\textbf{x})^*|^2 dA.
    \label{eqn:eq2}
\end{equation}
In our case, $\textbf{E}_d(\textbf{x})$ the electric field near the metasurface when LCP or RCP light is incident. $\textbf{E}_d(\textbf{x})$ is a LCP or RCP plane wave with a constant amplitude that propagates in the $-z$ direction. $\mathcal{M}$ is a monitor plane in the glass substrate and perpendicular to the $z$-axis, and * indicates a complex conjugate operation. Consequently, $\mathcal{F}$ is directly proportional to the flux of the transmitted LCP or RCP light when the LCP or RCP light is incident. The derivative of $\mathcal{F}$ by the transmitted electric field is 
\begin{equation}
    \frac{\partial \mathcal{F}}{\partial \textbf{E}(\textbf{x})} = \int_\mathcal{M}\textbf{E}_d(\textbf{x})^*\left[\textbf{E}(\textbf{x})^* \cdot \textbf{E}_d(\textbf{x})\right] dA.
    \label{eqn:eq3}
\end{equation}
The variation of the electric field at point $\textbf{x}$, caused by the adjustment in the permittivity of the design space, can be expressed as 
\begin{equation}
    \delta\textbf{E}(\textbf{x}) = \overleftrightarrow{\textbf{G}}(\textbf{x}, \textbf{x}') \textbf{P}^{\text{ind}}(\textbf{x}') = \overleftrightarrow{\textbf{G}}(\textbf{x}, \textbf{x}')\delta\epsilon(\textbf{x}')\textbf{E}(\textbf{x}'),
    \label{eqn:eq4}
\end{equation}
where $\textbf{x}$ and $\textbf{x}'$ indicate the positions in the monitor and the design space, respectively. In addition, $\textbf{P}^{\text{ind}}(\textbf{x}')$ indicates the polarization density,  which is induced by the variation of the dielectric constant $\delta\epsilon(\textbf{x}')$, and $\overleftrightarrow{\textbf{G}}(\textbf{x}, \textbf{x}')$ is a Green’s function which represents the electric field at the point $\textbf{x}$ generated by the unit dipole at the point $\textbf{x}'$. The variation of $\mathcal{F}$ is $\delta \mathcal{F} = \frac{\partial \mathcal{F}}{\partial \textbf{E}} \delta \textbf{E}+ \frac{\partial \mathcal{F}}{\partial \textbf{E}^*} \delta \textbf{E}^*$. The adjoint field $\textbf{E}_{\text{adj}}$ can be expressed as
\begin{equation}
\textbf{E}_{\text{adj}}(\textbf{x}') = \int_\mathcal{M}  \overleftrightarrow{\textbf{G}}(\textbf{x}, \textbf{x}') \cdot \left[ \textbf{E}_d(\textbf{x})^* (\textbf{E}(\textbf{x})^*\cdot\textbf{E}_d(\textbf{x}))\right]dA.
    \label{eqn:eq5}
\end{equation}
The adjoint field can be obtained by setting electric dipoles at the monitor plane with the direction and amplitude of $\textbf{E}_d^* \left(\textbf{E}^*\cdot\textbf{E}_d\right)$ because of the Lorentz reciprocity.~\cite{miller2012photonic} Finally, the gradient of $\mathcal{F}$ with respect to the permittivity at the point $\textbf{x}'$ can be expressed as 
\begin{equation}
    \frac{\partial\mathcal{F}}{\partial\epsilon(\textbf{x}')} = 2\text{Re}\left[ \textbf{E}(\textbf{x}') \cdot \textbf{E}_{\text{adj}}(\textbf{x}')\right].
    \label{eqn:eq6}
\end{equation}
To effectively modulate the geometric phase of the transmitted light, $T_{rr}=|t_{rr}|^2$, $T_{lr}=|t_{lr}|^2$, and $T_{ll}=|t_{ll}|^2$ should be minimized. At the same time, $T_{rl}=|t_{rl}|^2$ is required to be maximized, which is the value related to the efficiency of the metasurface. Therefore, we set the total FOM to maximize as a linear combination of partial FOMs, $\mathcal{F}_{rl}-\left(\mathcal{F}_{rr}+\mathcal{F}_{lr}+\mathcal{F}_{ll}\right)$, where the first and second subscripts of $\mathcal{F}$ refer to the circular polarization of the target field and the incidence in the forward field, respectively. The FOM is calculated at the wavelength of 680 nm. In addition, the forward and adjoint fields are simulated using the finite-difference time-domain (FDTD) method using a freely available software package, MEEP.~\cite{oskooi2010meep, lin2019topology}

\begin{figure}
    \centering
    \includegraphics[width=1.0\linewidth]{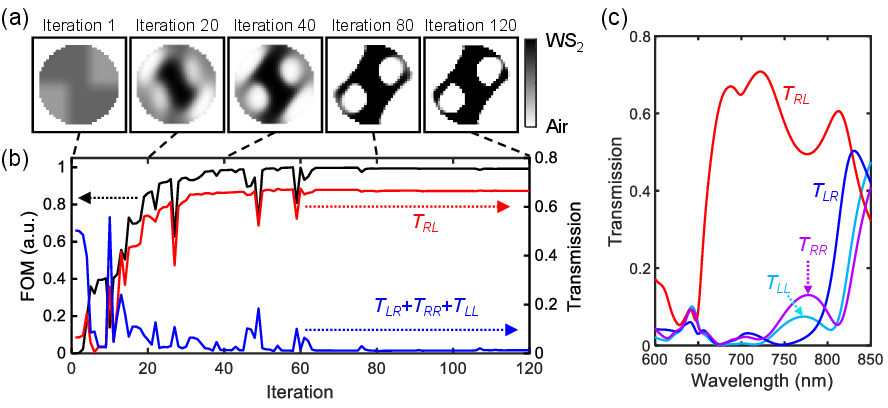}
    \caption{Inverse design procedure for optimising the WS$_2$ resonator. (a) The evolution of the design parameters is displayed with the iteration numbers. The colors black and white represent the permittivities of WS$_2$ and air, respectively, and the grey color indicates permittivity between those of air and WS$_2$. (b) The evolution of the figure of merit (FOM) and transmission in relation to the number of iterations. $T_{RL}$ and $T_{LL}$ indicate the transmitted light intensity of the RCP ($R$ in subscript) and LCP ($L$) light when LCP light is incident, respectively. Likewise, $T_{RR}$ and $T_{LR}$ refer to the transmission of the RCP and LCP light, when RCP light is incident. The red line shows $T_{RL}$, and the blue line represents the sum of three transmissions $T_{LR}+T_{RR}+T_{LL}$ at $\lambda$=680 nm. (c) The transmission spectra of the chiral resonator array depending on the incident and transmitted polarization, showing extremely small $T_{LR}$, $T_{RR}$, and $T_{LL}$ at $\lambda$=680 nm.
    }
    \label{fig:fig3}
\end{figure}

We applied additional design constraints and filters for the manufacturable design and effective geometric phase modulation. As shown in Fig. 3(a), the material outside of the design domain, a circle with 330 nm diameter, is fixed to the permittivity of air to reduce the interaction between neighboring resonators. Furthermore, C$_2$ symmetry was applied for full phase coverage, e.g., 0 to 2$\pi$. Moreover, a subpixel smoothing filter is applied to remove unfabricable fine structures.~\cite{lin2019topology} Binarization weights are used to make the design parameters converge to the permittivities of WS$_2$ and air.~\cite{bae2023inverse} Figure 3(b) shows the evolution of the total FOM and the transmissions with the increasing number of iterations. The FOM rapidly increases until it is saturated in about after the 40th iteration. After saturation, the binarization weights become dominant than the weights from the gradient of the FOM, pushing design parameters to converge to the permittivity of WS$_2$ and air. As the FOM increases, $T_{RL}$ increases while $T_{RR}$, $T_{LR}$, and $T_{LL}$ decrease. This indicates that the gradient of the FOM effectively enlarges $T_{RL}$ while suppressing the other transmissions. After the 120th iteration, the design parameters are fully binarized to the permittivities of either WS$_2$ or air. The optimized structure is shown in Fig.~\ref{fig:fig1}(c). Figure~\ref{fig:fig3}(c) shows the transmission spectra of the chiral resonator array depending on the incident and transmitted polarization. $T_{RR}$, $T_{RL}$, $T_{LL}$ show extremely low values ($<$0.006) compared to $T_{RL}$ ($>$0.65) at $\lambda$=680 nm.

\begin{figure}
    \centering
    \includegraphics[width=1.0\linewidth]{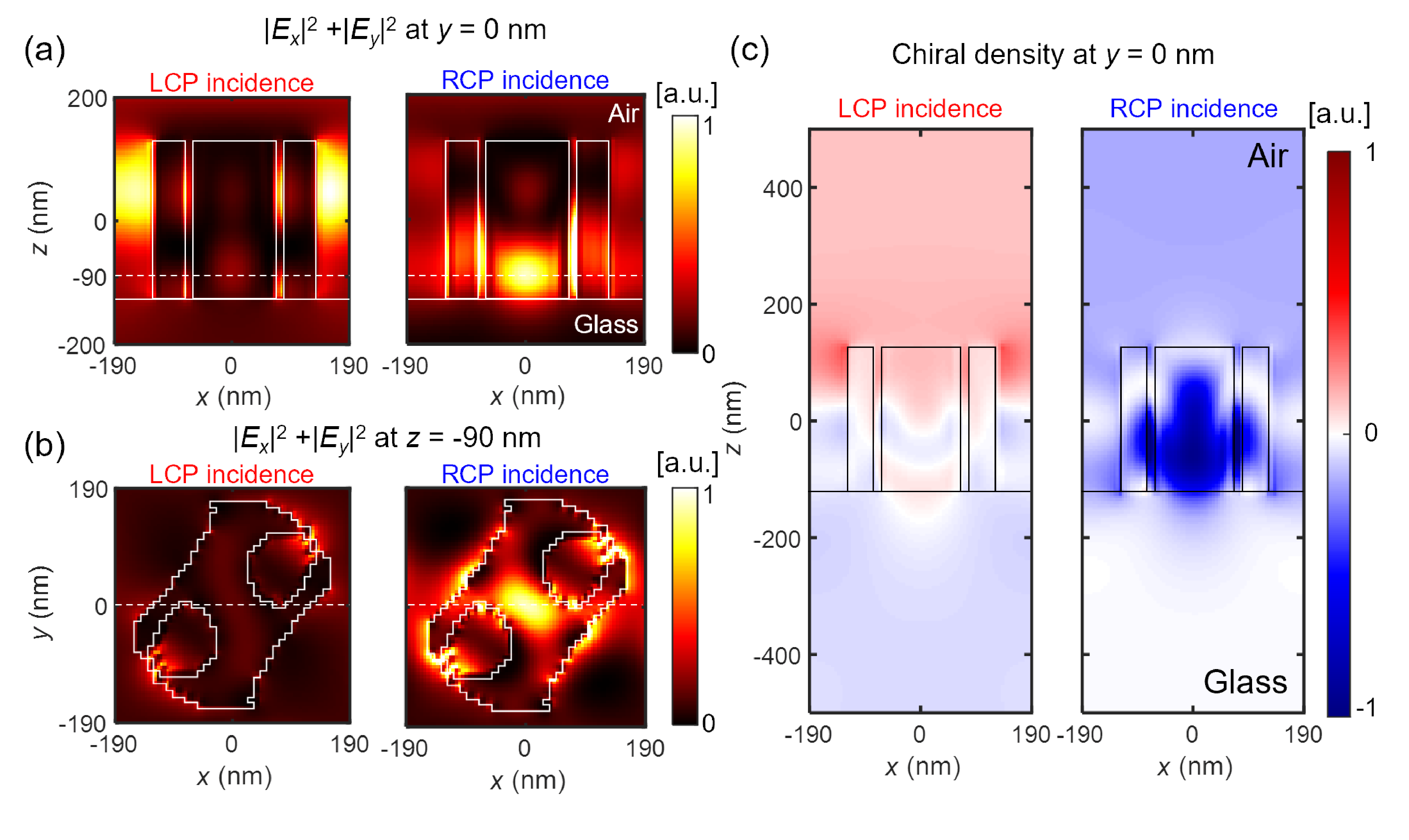}
    \caption{(a) Electric field intensities on the $xz$-plane under LCP and RCP incidences at $\lambda$=680 nm. The incident light propagates from top to bottom. Under LCP incidence, the strongest electric field intensity appears outside the WS$_2$ resonator. On the other hand, under RCP incidence, the strongest electric field appears inside the WS$_2$ resonator. (b) The electric near field intensities in the $xy$-plane at $z=-90$ nm under the LCP and RCP incidences at $\lambda$=680 nm. The RCP incidence creates highly concentrated fields inside the WS$_2$ resonator, which implicate high absorption compared to the LCP incidence. (c) Optical chiral densities under LCP and RCP incidences. The positive or negative chiral density indicates whether the chirality of LCP or RCP light is dominant at each point, respectively. 
    }
    \label{fig:fig4}
\end{figure}

Figures~\ref{fig:fig4}(a) and (b) present the electric field distributions under LCP and RCP incidences in $xy$-plane and $xz$-plane. Both LCP and RCP incidences experience resonant behavior, however, the distribution of the maximum electric field intensities differs with the incident polarization. Under LCP incidence, the lobe is mostly outside the chiral resonator, while it is mostly distributed inside the resonator for RCP incidence, causing comparatively large absorption and low transmission. The notable distinction in the electric field distribution, resulting in the large variation in absorption, may account for the large extinction ratio of the chiral resonator in this work. The optical chiral density~\cite{gryb2023two, tang2010optical} distribution is displayed in Fig.~\ref{fig:fig4}(c) to visualize the polarization conversion process induced by the chiral resonator. The optical chiral density is calculated from the equation,$-\frac{\epsilon_0\omega}{2}\text{Im}(\textbf{E}^*\cdot\textbf{B})$, and normalised by its maximum absolute value. Here, $\epsilon_0$ represents the vacuum permittivity, $\omega$ is the angular frequency, and $\textbf{B}$ is the magnetic field. The positive and negative chiral density indicate left-handed or right-handed chirality, which correspond to the chiralities of LCP and RCP lights, respectively. The magnitude of chiral density present a chiral intensity, which is directly proportional with the intensity ($\textbf{E}^*\times\textbf{B}$) when electric and magnetic fields are orthogonal. Under LCP incidence, positive and negative chiral densities are observed near the resonator, while only the negative chiral density is found at the bottom. This implies that LCP incidence is effectively converted to RCP light by the resonator. In the meantime, under RCP incidence, there exists no red color and the bottom shows white. This indicates that chiral handedness conversion is negligible, and the chirality is not transmitted.
\begin{figure}
    \centering
    \includegraphics[width=1.0\linewidth]{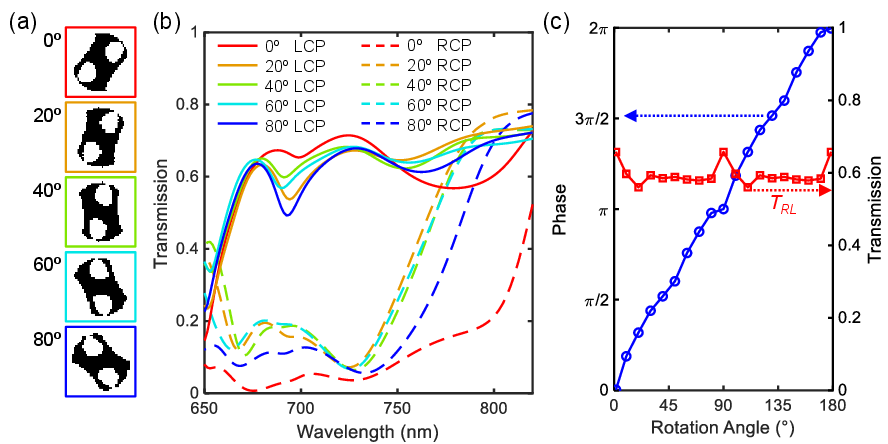}
    \caption{(a) The structures of the chiral resonators with various orientation angles. (b) Transmission spectra for a chiral resonator array with various orientation angles for the LCP (solid lines) and RCP (dashed lines) incidences. The resonator arrays show high circular dichroism with all orientation angles. (c) The transmittance $T_{RL}$ and the relative phases of the RCP light under the LCP incidence in the chiral resonator arrays with various orientation angles at $\lambda$=680 nm.}
    \label{fig:fig5}
\end{figure}

To verify the robustness of the chiral resonator, we tested the optical response with various orientation angles from \ang{0} to \ang{80} as shown in Fig.~\ref{fig:fig5}(a). Figure~\ref{fig:fig5}(b) shows transmission spectra of the chiral resonator arrays with varying orientation angles under RCP (dashed lines) and LCP (solid lines) incident light. 

They show strong circular dichroism across the wide range of the wavelengths. Figure~\ref{fig:fig5}(c) shows the relative phase of the transmitted RCP light under the LCP incidence (phase of $t_{rl}$) and transmission $T_{RL}$ at $\lambda$=680 nm. The transmission shows a lower bound of 0.56 over the desired bandwidth. Furthermore, the phase almost linearly increases with the rotation angle covering from 0 to 2$\pi$. It implies that we achieve a precise phase modulation using WS$_2$.

\section{Conclusion}
In this work, we have demonstrated an inversely designed WS$_2$ chiral metasurface. We utilized WS$_2$ as a platform material of the meta-atoms because of its significantly high refractive index and relatively low absorbance at the visible range, which provide great light tunability in an ultrathin layer. We also utilized the inverse design technique based on adjoint optimization to maximize cross-polarized transmission under LCP incidence while blocking RCP incidence. The metasurface exhibits consistently high CD ($>$0.5) across a wide range of wavelengths ($>$150 nm) and an extremely high extinction ratio (19.6dB). Moreover, we have demonstrated geometric phase modulation with the chiral resonator arrays with different orientation angles. They maintained high CD ($>$0.5) and transmission ($>$0.56) across all the orientation angles, demonstrating the capability for the geometric phase modulation. This chiral metasurface can realize ultra-compact geometric-phase-based metasurfaces by eliminating optical elements like QWPs and linear polarizers. Furthermore, it may open a new way to advance applications using the chirality of light, such as chiral-sensitive imaging and display technology and platforms of optical spin-orbit interactions.

\printbibliography
\end{document}